\documentclass[12pt]{iopart}
\usepackage[dvips]{graphicx}
\begin{document}

\title[Records in a changing world]{Records in a changing world}

\author{Joachim Krug}

\address{Institut 
f\"ur Theoretische Physik, Universit\"at zu K\"oln, 50937 K\"oln, Germany}
\ead{krug@thp.uni-koeln.de}
\begin{abstract}
In the context of this paper, a record is an entry in a sequence
of random variables (RV's) that is larger or smaller than all previous entries.
After a brief review of the classic theory of records, which is largely
restricted to sequences of independent and identically distributed (i.i.d.)
RV's, new results for sequences of independent RV's with distributions
that broaden or sharpen with time are presented. In particular, we show
that when the width of the distribution grows as a power law in time
$n$, the mean number of records is asymptotically of order $\ln n$ for
distributions with a power law tail (the \textit{Fr\'echet class} of extremal
value statistics), of order $(\ln n)^2$ for distributions of exponential
type (\textit{Gumbel class}), and of order $n^{1/(\nu+1)}$ for distributions
of bounded support (\textit{Weibull class}), where the exponent $\nu$
describes the behaviour of the distribution
at the upper (or lower) boundary. Simulations 
are presented which indicate that, in contrast to the i.i.d. case, the
sequence of record breaking events is correlated in such a way that the
variance of the number of records is asymptotically smaller than the mean. 
\end{abstract}

\section{Introduction}
\label{Sec:Intro}
A record is an entry in a discrete time series that is larger (\textit{upper
record}) or smaller (\textit{lower record}) than all previous entries. Thus,
records are extreme values that are defined not relative to a fixed threshold,
but relative to all preceding events that have occurred since the beginning
of the process. Statistical data in areas like meteorology 
\cite{Hoyt81,Bassett92,Benestad03,Redner06}, hydrology
\cite{Matalas97,Vogel01} and athletics \cite{Gembris02}
are naturally represented in terms of records. Records play an 
important role in the public perception of issues like anthropogenic
climate change and 
natural disasters such as floods and earthquakes, and they are an integral
part of popular culture. Indeed, the \textit{Guinness Book of Records}, 
first published in 1955, is the world's most sold copyrighted book. 

The mathematical theory of records was initiated more than 50 years ago 
\cite{Chandler52}, and it is now a mature subfield of probability theory
and statistics;
see \cite{Glick78,Nevzorov87,Arnold98,Nevzorov01} for reviews and
\cite{Schmittmann99} for an elementary introduction. Most of this
work has been devoted to the case when the time series under consideration
consists of independent, identically distributed (i.i.d.) random variables
(RV's). For the following discussion, it will be useful to distinguish between
the \textit{record times} at which the current record
is broken and replaced by a new one, and the associated
\textit{record values}. One of the key results of record theory is that
the statistical properties of record times
for real-valued i.i.d. RV's are completely 
independent of the underlying distribution. To illustrate the origin of this
universality, we recall the basic observation that the probability 
$P_n$ for a record to occur in the $n$'th time step (the \textit{record
rate}) is given by 
\begin{equation}
\label{rate}
P_n = \frac{1}{n}
\end{equation}
for i.i.d. RV's, because each of the $n$ first entries $X_1,...,X_n$,
including the last, is equally likely to be the largest or smallest.      
The mean number of records up to time $n$, $\overline{R_n}$, is therefore 
given by the harmonic series 
\begin{equation}
\label{harmonic} 
\overline{R_n} = \sum_{k=1}^n P_k = 
\sum_{k=1}^n \frac{1}{k} \approx \ln n + 
\gamma +{\cal{O}}(1/n)  \;\;\;
{\mathrm{for}} \;\;\; n \to \infty,
\end{equation}
with $\gamma \approx 0.5772156649...$. Further 
considerations along the same lines lead to a remarkably complete
characterization of record times, which will be briefly reviewed below in 
section \ref{Sec:iid}. The universality of record times can be 
exploited in statistical tests of the i.i.d. property of a given sequence
of variables, without the need for any hypothesis about the underlying 
distribution \cite{Glick78}. 
By contrast, distributions of record values fall into three
distinct universality classes, which are largely analogous to the well-known
asymptotic laws of extreme value statistics for distributions with
exponential-like tails (\textit{Gumbel}), bounded support (\textit{Weibull})
and power law tails (\textit{Fr\'echet}), respectively 
\cite{Galambos87,Sornette00}.

The decay of the record rate (\ref{rate}) with increasing $n$ implies that
the record breaking events form a non-stationary time series 
with unusual statistical properties, which will be further discussed below
in section \ref{Sec:iid}. \textit{Record dynamics} has therefore been proposed
as a paradigm for the non-stationary temporal behaviour of diverse
complex systems ranging from the low-temperature relaxation of 
spin glasses to the co-evolution of biological populations  
\cite{Sibani93,Sibani03,Anderson04}. In fact, records
appear naturally in the theory of biological adaptation, because any
evolutionary innovation that successfully spreads in a population
must be a record, in the sense that it accomplishes some task encountered
by the organism in a way that is superior to all previously existing
solutions. Consequently the statistics of records and 
extremes has been invoked to understand the distribution of fitness
increments in adaptive processes \cite{Gillespie91,Orr05} 
as well as the timing of
adaptive events \cite{Kauffman87,Sibani98,Krug03,Krug05,Jain05,Sire06}. 
In the biological context the universality of record time statistics
is particularly attractive, because genotypical fitness is a somewhat 
elusive notion that is hard to quantify in terms of explicit probability 
distributions. 

Surprisingly few result on record statistics are known that go beyond the
standard setting of i.i.d. RV's, and thus consider correlated and/or 
non-identically distributed RV's. In the present article we focus
exclusively on the latter issue, while maintaining the independence among
the entries in the sequence. A simple example of this type was introduced
by Yang in an attempt to explain the frequency of occurrence
of Olympic records, which is much higher than would be expected on the 
basis of the i.i.d. theory \cite{Yang75}. In his model a specified number
of i.i.d. RV's become available simultaneously in each time step, 
corresponding, in the athletic context, to a variable (growing) population
from which the contenders are drawn. Much of the standard theory can be 
extended to this case \cite{Nevzorov87,Nevzorov01} 
(see section \ref{Sec:Growing} for a brief review). In particular, one finds
that the record rate becomes asymptotically constant for exponentially
growing populations. 
An application of Yang's model to evolutionary searches in the space
of genotypic sequences can be found in \cite{Krug05,Jain05}.
A second line of research has addressed the case of 
sequences with a linear trend, in which the $n$'th entry is of the form
\begin{equation}
\label{trend}
X_n = Y_n + cn
\end{equation}
with i.i.d. RV's $Y_n$ and $c > 0$ \cite{Ballerini85,Ballerini87,Borovkov99}.
Also in this case the record rate becomes asymptotically constant,
see section \ref{Sec:Growing} for details.

The effect of trends on the occurrence rate of records is a key issue
in the ongoing debate about the observable consequences
of global warming \cite{Hoyt81,Bassett92,Benestad03,Redner06,Matalas97}. 
In this context it has
been pointed out that climate \textit{variability} is presumably a 
more important factor in determining the frequency of extreme events than
averages \cite{Katz92}. It is therefore of considerable interest
to investigate the record statistics of sequences of uncorrelated 
RV's in which 
the \textit{shape} of the underlying probability distribution changes 
systematically with time. To initiate such an investigation is the goal
of the present paper. Throughout we assume that the probability density 
$p_n(X)$ of the $n$'th entry $X_n$ is of the form
\begin{equation}
\label{shape}
p_n(X) = \lambda_n \; \Pi(\lambda_n X)
\end{equation}
where $\Pi(X)$ is a fixed normalized distribution and the $\lambda_n$ 
usually have a power-law time dependence
\begin{equation}
\label{power}
\lambda_n = \lambda_0 \; n^{-\alpha},
\end{equation}
so that $\alpha > 0$ ($\alpha < 0$) 
corresponds to a broadening (sharpening) distribution.

After a brief review of a few important classic results of the 
theory of records in section \ref{Sec:Classic}, our new results for
non-indentically distributed random variables will be presented in 
section \ref{Sec:record}. We focus on the asymptotic behaviour
of the record rate $P_n$ and the mean number of records
$\overline{R_n}$. Preliminary numerical results for the variance of 
the number of records are reported in section \ref{Sec:Numerics}, but
a more complete characterization of record times and
record values is left for future work. Finally, some concluding remarks
are offered in section \ref{Sec:Discussion}.

\section{Brief survey of classic results}
\label{Sec:Classic}

Given the distributions $p_k(X)$ of the entries $X_k$ in a sequence of 
independent RV's, the probability $P_n$ that the $n$'th entry
is an upper record is equal to the probability that $X_n > X_k$ for all
$k < n$. Hence\footnote{Here and in the following limits of integration are
omitted whenever the domain of integration is understood to comprise the
entire support of the probability distribution.} 
\begin{equation}
\label{upper}
P_n = \int dX_n \; p_n(X_n) \prod_{k=1}^{n-1} q_k(X_n),
\end{equation}
where  
\begin{equation}
\label{cumulative}
q_k(X) = \int^{X} dx \; p_k(x)
\end{equation}
is the cumulative distribution of $X_k$. Similarly the probability that
$X_n$ is a lower record reads
\begin{equation}
\label{lower}
P_n^\ast = \int dX_n \; p_n(X_n) \prod_{k=1}^{n-1} [1 - q_k(X_n)].
\end{equation}
Equations (\ref{upper}) and (\ref{lower}) form the basis for most
of what follows.

\subsection{Records from i.i.d. random variables}
\label{Sec:iid}

For i.i.d. RV's the integral (\ref{upper}) can be performed by noting 
that $p_k, q_k \equiv p,q$ and $dq = p \; dX$, which yields the universal
result (\ref{rate}). To arrive at a characterization of the record time
process beyond the mean number of records $\overline{R_n}$ we introduce
the \textit{record indicator variables} $I_n$, which take the value
$I_n = 1$ iff $X_n$ is a record, and $I_n = 0$ else. It turns out that
the $I_n$ are independent \cite{Glick78,Arnold98}, 
and hence they form a Bernoulli process with success probability $P_n$.
To see why this is so, consider the two-point correlation function 
$\overline{I_i I_j}$ and assume that $j > i$. Then the key idea is that
the right hand side of 
\begin{equation}
\label{ij1}
\overline{I_i I_j} = {\mathrm{Prob}}[X_i = \max(X_1,...,X_i) \;\; 
{\mathrm{and}} \;\; X_j = \max(X_1,...,X_j)]
\end{equation}
can be split into independent events according to 
\begin{eqnarray}
\label{ij2} 
\fl
\overline{I_i I_j} = {\mathrm{Prob}}[X_i = \max(X_1,...,X_i)] \times
{\mathrm{Prob}}[X_j = \max(X_{i+1},...,X_j)] \times \nonumber \\
\times {\mathrm{Prob}}[\max(X_1,...,X_i) < \max(X_{i+1},...,X_j)].
\end{eqnarray}
Following the symmetry argument used to derive (\ref{rate}), the first
two factors are $1/i$ and $1/(j-i)$, respectively, and the third factor
can be written as 
\begin{equation}
\label{ij3}
{\mathrm{Prob}}[\max(X_1,...,X_j) \;\; 
{\mathrm{occurs}} \; {\mathrm{in}} \;\; \{X_{i+1},...,X_j\}] = 
\frac{j-i}{j}.
\end{equation}
We conclude that 
\begin{equation}
\label{ij4} 
\overline{I_i I_j} = \frac{1}{i} \; \frac{1}{j-i} \;
\frac{j-i}{j} = \frac{1}{i} \;
\frac{1}{j} = P_i P_j = (\overline{I_i}) (\overline{I_j}).
\end{equation} 
Higher order correlations can be shown to factorize in the same way.
The number $R_n$ of records up to time $n$
can then be expressed in terms of the indicator variables as 
\begin{equation}
\label{PnIk}
R_n = \sum_{k=1}^n I_k,
\end{equation}
and the variance of $R_n$ is 
\begin{equation}
\fl
\label{Rn_variance}
\overline{(R_n - \overline{R_n})^2} = \sum_{k=1}^n (P_k - P_k^2) 
= \sum_{k=1}^n \left( \frac{1}{k} - \frac{1}{k^2} \right) \approx
\ln n + \gamma - \pi^2/6 + {\cal{O}}(1/n)
\end{equation}
for $n \to \infty$.
The \textit{index of dispersion} $\rho_n$
of the record time process \cite{Gillespie91,Cox80}, defined as
the ratio of the variance to the mean
\begin{equation}
\label{rho}
\rho_n = \frac{\overline{(R_n - \overline{R_n})^2}}{\overline{
R_n}}
\end{equation} 
thus tends to unity, and
the distribution of the $R_n$ becomes Poissonian
with mean $\ln n$ for large $n$. The 
record times form a \textit{log-Poisson} process 
\cite{Sibani93,Sibani03,Sibani98}. 

A second useful observation concerns the ratios between consecutive record
times. Let $t_m$ denote the time of the $m$'th record, with $t_1 = 1$
by convention. Repeating the symmetry argument used to derive (\ref{rate}),
we expect that given $t_m$, the preceding $(m-1)$'th record occurs with
equal probability anywhere in the interval $[1, t_{m}]$. This is not
quite correct, because the previous $m-2$ records also have to be 
accomodated, but since $m \sim \ln(t_m)$ this is a small correction which
can be neglected for large $m$. It is therefore plausible (and can be 
proved \cite{Tata69}) that the ratio $t_{m-1}/t_m$ tends to a uniformly
distributed RV $u_m \in [0,1]$ for large $m$. Moreover the $u_m$ become
independent in this limit \cite{Shorrock72}. This allows us to highlight
a peculiar property of the sequence of record breaking events: The 
expected value of $t_{m-1}$, given $t_m$, is 
\begin{equation}
\label{backward}
\overline{t_{m-1}} \vert_{t_m} = t_m \int_0^1 du \; u = \frac{1}{2} t_m,
\end{equation}
but the reverse conditioning yields an \textit{infinite} expectation, 
because 
\begin{equation}
\label{forward}
\overline{t_m} \vert_{t_{m-1}} = t_{m-1} \int_0^1 du \; u^{-1} = \infty.
\end{equation}
In this sense, the occurrence of 
records can be predicted only with hindsight, but not forward in time.

\subsection{Growing and improving populations}
\label{Sec:Growing}

In the model for growing populations introduced by Yang \cite{Yang75} and 
elaborated by Nevzorov \cite{Nevzorov87}, a number $N_n$ of of i.i.d. 
RV's becomes available simultaneously at time $n$. The symmetry argument
in section \ref{Sec:Intro} is easily extended to this case: Because of the 
i.i.d. property, the probability that there is a record among the $N_n$
newly generated RV's is equal to the ratio of $N_n$ to the total number
of RV's that have appeared up to time $n$, and hence
\begin{equation}
\label{rate_Yang}
P_n = \frac{N_n}{\sum_{k=1}^n N_k}.
\end{equation}
The independence of the record indicator variables $Y_n$ introduced above 
in section \ref{Sec:iid} continues to hold
\cite{Nevzorov87,Jain05}, so again the sequence of record
breaking events is a Bernouilli process with success probability $P_n$.

To give a simple example for the consequences of
(\ref{rate_Yang}), suppose the $N_n$ grow exponentially as  
as $a^n$ with $a > 1$. This could model a sequence of athletic competitions
in an exponentially growing population, where each athlete is assumed to 
be able to participate only in one event \cite{Yang75}. Then 
the evaluation of (\ref{rate_Yang}) yields  
\begin{equation}
\label{rate_exp}
P_n = \frac{a^n(a-1)}{a(a^n-1)} \to \frac{a-1}{a} \;\;\;
{\mathrm{for}} \;\;\; n \to \infty,
\end{equation}
and the distribution of inter-record times $t_m - t_{m-1}$ is geometric.
In his analysis of Olympic records Yang estimated a growth factor of 
$a \approx 1.08$ for the four-year period between two games, and concluded
that this growth rate was insufficient to explain the observed high frequency
of records.

Motivated by this outcome, 
Ballerini and Resnick \cite{Ballerini85} considered
a model of \textit{improving} populations, where the sequence of RV's
displays a linear drift according to (\ref{trend}). They showed that the
record rate tends to an asymptotic limit $P(c)$ given by
\begin{equation}
\label{limitrate}
P(c) = \lim_{n \to \infty} P_n = 
\int dy \; p(y) \; G_\infty(y),
\end{equation}
where $p(Y)$ is the probability density of the i.i.d. RV's $Y_k$ in 
(\ref{trend}) and 
\begin{equation}
\label{Ginf}
\fl
G_\infty(y) = \lim_{n \to \infty} {\mathrm{Prob}}[Y_k - ck \leq y \;\;
{\mathrm{for}} \; {\mathrm{all}} \; k = 1,...,n-1]
= \lim_{n \to \infty} \prod_{k=1}^{n-1} q(y + ck),
\end{equation}
with $q(Y) = \int^Y dz \; p(z)$. The function $P(c)$ has the obvious limits
$P(0) = 0$ and $\lim_{c \to \infty} P(c) = 1$, but the explicit evaluation
is generally difficult. A simple expression is obtained when     
$q(Y)$ is of Gumbel form, $q(Y) = \exp[-e^{-Y/b}]$, which yields
$P(c) = 1 - e^{-c/b}$. For further details on the model (\ref{trend})
and applications to athletic data we refer to 
\cite{Ballerini85,Ballerini87,Borovkov99}. Results for specific distributions
and an application to global warming can be found in \cite{Redner06}.

\section{Records in sequences with increasing or decreasing variance}
\label{Sec:record}

In this section we want to evaluate the record rates (\ref{upper}) and 
(\ref{lower}) for distributions of the general form (\ref{shape}).
Introducing the cumulative distribution corresponding to $\Pi(X)$,
\begin{equation}
\label{Q}
Q(X) = \int^X dx \; \pi(x),
\end{equation}
the record rates of interest can be written as
\begin{equation}
\label{upper1}
P_n = \int dz \; \Pi(z) \prod_{k=1}^{n-1} Q(z \lambda_k/\lambda_n)
\end{equation}
and
\begin{equation}
\label{lower1}
P_n^\ast = \int dz \; \Pi(z) \prod_{k=1}^{n-1} [1 - Q(z \lambda_k/\lambda_n)],
\end{equation}
which makes clear the obvious fact that the overall scale of the 
$\lambda_n$'s is without importance.

\subsection{Simple cases}

In some special cases the record rates can be evaluated exactly for arbitrary
choices of the $\lambda_n$'s. For example, for the exponential
distribution 
\begin{equation}
\label{exponential}
\Pi(X) = e^{-X}, \;\;\; X \geq 0
\end{equation}
we have $Q(X) = 1 - e^{-X}$, and the evaluation of the lower record rate
(\ref{lower1}) yields
\begin{equation}
\label{lower_exp}
P_n^\ast = \frac{\lambda_n}{\sum_{k = 1}^n \lambda_k}.
\end{equation}
Inserting the power law behaviour (\ref{power}) we see that the 
denominator converges to the Riemann zeta function $\zeta(\alpha)$
for $\alpha > 1$, so that $P_n^\ast \to n^{-\alpha}/\zeta(\alpha)$
for large $n$, and the expected number of lower 
records 
\begin{equation}
\label{lower_number}
\overline{R_n^\ast} = \sum_{k=1}^n P_k^\ast
\end{equation} 
remains finite for $n \to \infty$. For $\alpha < 1$ we have instead that
$P^\ast_n \approx (1-\alpha)/n$ for large $n$, and hence  
\begin{equation}
\label{lower_small}
\overline{R_n^\ast} \approx (1 - \alpha) \ln n
\end{equation}
asymptotically. As would be intuitively expected, the occurrence of lower
records is enhanced for sharpening distributions ($\alpha < 0$) 
and suppressed for broadening distributions ($\alpha > 0$). Finally, 
in the borderline case $\alpha = 1$ we find
\begin{equation}
\label{borderline}
\overline{R_n^\ast} \approx \ln(\ln(n)),
\end{equation}
which is our first example of a nontrivial asymptotic law that differs
qualitatively from the i.i.d. result (\ref{harmonic}). 

A simple explicit expression for the upper record rate $P_n$ can be obtained
for the uniform distribution characterized by 
\begin{equation}
\label{uniform}
Q(X) = X \;\;\; {\mathrm{for}} \;\;\; 0 \leq X \leq 1
\end{equation}
when the $\lambda_k$ are increasing, in the sense that $\lambda_k/\lambda_n
< 1$ for all $k < n$, i.e. for the case of a sharpening uniform distribution.
Then the arguments of $Q$ on the right hand side of (\ref{upper1}) are all
less than unity, and direct integration yields
\begin{equation}
\label{uniform1}
P_n = \frac{1}{n} \prod_{k=1}^{n-1} \frac{\lambda_k}{\lambda_n}.
\end{equation}
Inserting the power law form (\ref{power}) with $\alpha < 0$ one finds
that the record rate decays exponentially as $P_n \sim e^{\alpha n}$,
and hence the asymptotic number of records is finite for all $\alpha < 0$.

\subsection{Asymptotics of the mean number of records}

In this section we focus on broadening distributions, $\alpha > 0$, and
evaluate the upper record rate (\ref{upper1}) asymptotically
for representatives
of all three universality classes of extreme value statistics. 
The starting point is to replace the product on the right hand side 
of (\ref{upper1}) by the exponential of a sum of logarithms, and to 
replace the latter by an integral. It then follows that the asymptotic
behaviour of the record rate is given by 
\begin{equation}
\label{asym}
P_n \approx \int dz \; \Pi(z) e^{n g_\alpha(z)} = 
\int_0^1 dQ \; e^{n g_\alpha(z(Q))}.
\end{equation}
The second representation will prove to be useful in the final 
evaluation of $P_n$. Note that $z$ can always be expressed in terms of $Q$
because $dQ/dz = \Pi \geq 0$. The function $g_\alpha$ is given by 
\begin{equation}
\label{galpha}
g_\alpha(z) = \int_0^1 du \; \ln Q(z/u^\alpha) \approx
- \int_0^1 du \; (1 - Q(z/u^\alpha)), 
\end{equation} 
where in the second step it has been used that the integral 
in (\ref{asym}) is dominated for large $n$ by the region where
$g_\alpha \to 0$ and $Q \to 1$. It is therefore clear that the 
asymptotic behaviour of the record rate depends only on the tail of
$Q$, and hence universality in the sense of standard extreme value
statistics should apply. 

\subsubsection{Fr\'echet class}

The evaluation of (\ref{galpha}) is straightforward for 
the Fr\'echet class of distributions with power law tails.
We set
\begin{equation}
\label{Frechet}
Q(X) = 1 - X^{-\mu}, \;\;\;\; X \geq 1
\end{equation}
and obtain 
\begin{equation}
\label{ga_Frechet}
g_\alpha(z) \approx -(1 + \alpha \mu)^{-1} z^{-\mu} = 
- (1 + \alpha \mu)^{-1} (1 - Q(z)).
\end{equation} 
Inserting this into (\ref{asym}) yields
\begin{equation}
\label{Pn_Frechet}
P_n \approx \int_0^1 dQ \; e^{-n(1 + \alpha \mu)^{-1} (1 - Q)}
\to \frac{1 + \alpha \mu}{n}
\end{equation}
for large $n$, and hence
\begin{equation}
\label{R_Frechet}
\overline{R_n} \approx (1 + \alpha \mu) \ln n.
\end{equation}
This result remains valid for negative $\alpha$ as long as 
$\alpha \mu > -1$. When $\alpha \mu < -1$ the evaluation of $g_\alpha$
shows that $P_n \sim n^{\alpha \mu}$ and thus the asymptotic number of 
records remains finite. 

\subsubsection{Gumbel class}

The Gumbel class comprises unbounded distributions whose tail decays
faster than a power law \cite{Galambos87,Sornette00}. 
A typical representative is
the exponentical distribution (\ref{exponential}) with $Q(X) = 1 - 
e^{-X}$. Evaluation of (\ref{galpha}) yields
\begin{equation}
\label{Gamma}
g_\alpha(z) \approx - \frac{z^{1/\alpha}}{\alpha} \int_z^\infty dv \;
v^{-(1 + 1/\alpha)} e^{-x} = 
- \frac{z^{1/\alpha}}{\alpha} \Gamma(-1/\alpha,z),
\end{equation}
where $\Gamma(-1/\alpha,z)$ denotes the incomplete gamma function.
For large $z$ we have \cite{Gradshteyn00} 
\begin{equation}
\label{Gamma2}
\Gamma(-1/\alpha,z) \approx z^{-(1 + 1/\alpha)} e^{-z},
\end{equation}
so that 
\begin{equation}
\label{galpha_exp}
g_\alpha(z) \approx - \frac{e^{-z}}{\alpha z} = 
\frac{1 - Q(z)}{\alpha \ln(1 - Q(z))},
\end{equation}
which yields
\begin{equation}
\label{Pn_exp}
P_n \approx \int_0^1 dQ \; \exp\left[\frac{n(1 - Q)}{\alpha \ln (1 - Q)}
\right] =  \int_0^1 dv \; \exp[-n v/(\alpha \ln(1/v))].  
\end{equation}
To further evaluate the integral we substitute $w = (n/\ln n) v$ and
obtain 
\begin{eqnarray}
\label{Pn_exp2}
P_n \approx \frac{\ln n}{n} \int_0^{n/\ln n} dw \; 
\exp \left[- \frac{w \ln n}{\alpha (\ln n - \ln (\ln n) - \ln w)} \right]
\to \nonumber \\
\to \frac{\ln n}{n} \int_0^\infty dw \; e^{-w/\alpha} = 
\frac{\alpha \ln n}{n}
\end{eqnarray}
for $n \to \infty$. Correspondingly the mean number of records grows as
\begin{equation}
\label{mean_exp}
\overline{R_n} \approx \frac{\alpha}{2} (\ln n)^2. 
\end{equation}
 
A second important representative of the Gumbel class is the Gaussian
(normal) distribution, for which 
\begin{equation}
\label{QGauss}
Q(X) \approx 1 - \frac{1}{2 \sqrt{\pi}} \frac{e^{-X^2}}{X}
\;\;\; {\mathrm{for}} \;\;\; X \to \infty.
\end{equation}
Proceeding as before, we find
\begin{equation}
\label{galpha_Gauss}
g_\alpha(z) \approx - \frac{z^{1/\alpha}}{4 \sqrt{\pi} \alpha} 
\Gamma(-1/2 -1/2 \alpha, z^2) \approx - \frac{1 - Q(z)}{2 \alpha z^2}
\approx \frac{1-Q}{2 \alpha \ln(1 - Q)},
\end{equation}
which becomes identical to (\ref{galpha_exp}) upon replacing $\alpha$
by $2 \alpha$. We conclude that $P_n \approx (2 \alpha \ln n)/n$
and $\overline{R_n} \approx 2 \alpha (\ln n)^2$ for the Gaussian case.
Although this does not constitute a strict proof, it strongly indicates
that the behaviour $\overline{R_n} \sim (\ln n)^2$ is \textit{universal} within
this class of probability distributions. 

\subsubsection{Weibull class}

As a representative of the Weibull class of distributions with finite
support we first consider the uniform distribution (\ref{uniform}). 
The integral on the right hand side of (\ref{galpha}) can then be 
evaluated without approximating $\ln Q$ by $-(1-Q)$, and one obtains
\begin{equation}
\label{galpha_uni}
g_\alpha(z) = \int_{z^{1/\alpha}}^1 du \; 
\ln \left(\frac{z}{u^\alpha} \right) = \ln z + \alpha(1 - z^{1/\alpha}).
\end{equation}
This is a negative monotonically increasing function which vanishes
quadratically in $1-z$ near $z = 1$, 
\begin{equation}
\label{quad}
g_\alpha(z) \approx - \frac{1}{2 \alpha} (1 - z)^2 = 
-\frac{1}{2 \alpha} (1 - Q)^2 \;\;\; 
{\mathrm{for}} \;\;\; z,Q \to 1.
\end{equation}
The evaluation of the record rate (\ref{asym}) then yields
\begin{equation}
\label{Pn_uni}
P_n \approx \int_0^1 dQ \; \exp[-n(1 - Q^2)/2 \alpha] \approx \sqrt{
\frac{\alpha \pi}{2 n}}
\end{equation}
for large $n$, and the number of records grows asymptotically as 
$\sqrt{n}$. The specific power is clearly related to the quadratic
behaviour of $g_\alpha$ near $z = 1$, which in turn reflects the
behaviour of $Q(X)$  near the upper boundary $X=1$. More generally
we may consider bounded distributions of the form
\begin{equation}
\label{bounded}
Q(X) = 1 - (1 - X)^\nu, \;\;\; 0 \leq X \leq 1,
\end{equation}
with $\nu > 0$ and the uniform case corresponding to $\nu = 1$. 
To extract the leading order behaviour of $g_\alpha$ for $z \to 1$
we write
\begin{eqnarray}
\label{galpha_bounded} 
g_\alpha(z) \approx - \int_{z^{1/\alpha}}^1 du \; (1 - z/u^\alpha)^\nu = 
-\frac{z^{1/\alpha}}{\alpha} \int_z^1 dv v^{-(1+1/\alpha)} (1 - v)^\nu
\approx \nonumber \\
\approx -\frac{1}{\alpha(\nu+1)} (1 - z)^{\nu + 1}
\end{eqnarray}
for $z \to 1$. Hence the record rate decays as $n^{-\nu/(\nu+1)}$ and 
the mean number of records grows as 
\begin{equation}
\label{Weibull_mean}
\overline{R_n} \approx \frac{\nu \Gamma(\nu/(\nu+1))}{(\nu+1)^{1+1/(\nu+1)}}
(\alpha^\nu n)^{1/(\nu+1)}.
\end{equation}

\subsection{Simulations}
\label{Sec:Numerics}

\begin{figure}
\centerline{\includegraphics[width=0.7\textwidth]{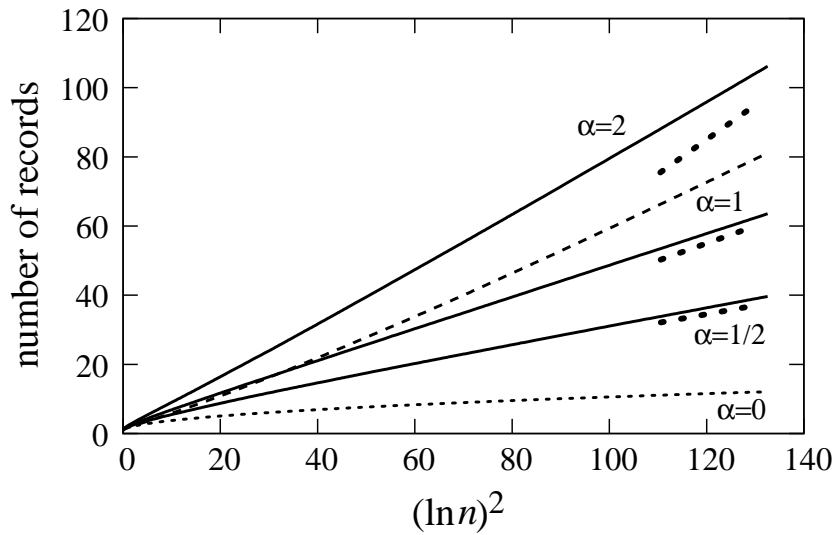}}
~\caption{Simulation results for the mean 
number of records for distributions of 
Gumbel type. Full lines show data obtained for the exponential distribution
with $\alpha = 2$, $\alpha = 1$ and $\alpha = 1/2$. The dashed line
shows data obtained for the Gaussian distribution and $\alpha = 1$.
The thin dotted line is the harmonic series (\ref{harmonic}) which applies
universally for $\alpha = 0$. The short bold dotted lines show the predicted
slope $\alpha/2$ for the exponential case and $\alpha$ in the Gaussian
case. All data were obtained from $10^4$ realizations of time series of
length $10^5$.}  
\label{Figure1}
\end{figure}

The asymptotic laws (\ref{R_Frechet}, \ref{mean_exp}, \ref{Weibull_mean})
were first discovered in simulations, and they have 
subsequently been numerically verified for a variety of parameter values.
As an example, we show in Figure \ref{Figure1} numerical data for the mean
number of records obtained for distributions in the Gumbel class.  
There are significant corrections to the asymptotic behaviour for the 
Gaussian distribution as well as for the exponentical distribution with
$\alpha = 2$. This is not surprising in view of the approximations used
in the derivation of (\ref{mean_exp}); for example, the 
last step in (\ref{Pn_exp2}) requires that $\ln n \gg \ln (\ln n)$
which is true only for enormously large values of $n$.

\begin{figure}
\centerline{\includegraphics[width=0.7\textwidth]{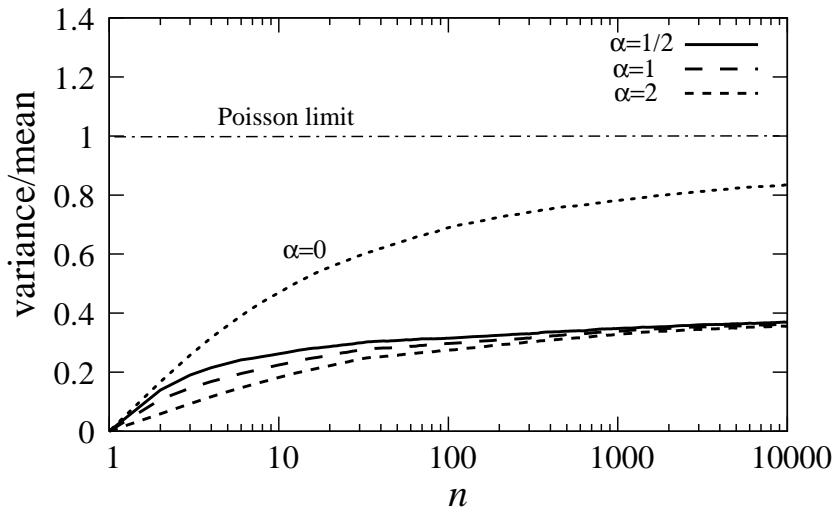}}
\caption{Simulation results for the ratio of the variance of the number
of records to the mean obtained using the uniform distribution with
$\alpha =0$, 1/2, 1 and 2. While the data for $\alpha = 0$ approach
the asymptotic Poisson limit of unity according to 
(\ref{Rn_variance}), the data for $\alpha > 0$
converge to a universal sub-Poissonian value. The data were obtained from
$10^5$ realizations of time series of length $10^4$.}
\label{Figure2}
\end{figure}

Simulations have also been used to investigate the occurrence of correlations
in the record time process for $\alpha > 0$. We have seen in 
section \ref{Sec:iid}
that the Poisson statistics of $R_n$ is a consequence of the fact that 
the record indicator variables $I_n$ are independent in the i.i.d. case. In 
particular, (\ref{Rn_variance}) 
shows that the variance of $R_n$ is asympotically 
equal to the mean whenever the $I_n$ are uncorrelated and the record
rate $P_n$ tends to zero for $n \to \infty$ in such a way 
that $\overline{R_n}$ diverges. As this is true for $\alpha > 0$ in all
cases that we have considered, the index of dispersion 
(\ref{rho}) can be used as a probe for correlations. The data displayed in 
Figure \ref{Figure2} clearly show that the asymptotic value of $\rho_n$
is less than unity and independent of $\alpha$ for the uniform distribution.
Similar results have been obtained for the exponential distribution,
whereas we find that
$\rho_n \to 1$ for the power law case. We conclude that, at least
in certain cases, the record time process becomes more regular 
than the log-Poisson process when
the underlying distribution broadens with time. 

\section{Summary and discussion}
\label{Sec:Discussion}

The main results of this paper are the asymptotic laws 
(\ref{R_Frechet}, \ref{mean_exp}, \ref{Weibull_mean}) for the mean number
of records in sequences of random variables drawn from broadening 
distributions. In all three cases the exponent $\alpha$ governing the time
dependence of the width of the distribution enters only in the prefactors 
and does not affect the functional form of the result. Comparing the three
cases, we see that the effect of the broadening 
on $\overline{R_n}$ is stronger the faster
the underlying distribution $\Pi(X)$ decays for large arguments: 
For fat-tailed power law distributions the number of records remains 
logarithmic, for exponential-like distributions it changes from $\ln n$
to $(\ln n)^2$, while for distributions with bounded support the 
logarithm speeds up to a power law in time. 

Apart from the presentation of new results, a secondary purpose of this
paper has been to advertise record dynamics as a
paradigm of non-stationary point processes with interesting
mathematical properties and wide-spread applications ranging from
fundamental issues in the dynamics of complex systems to 
the consequences of climatic change. In the present work we have combined
the intrinsic non-stationarity of record dynamics with an explicit
non-stationarity of the underlying sequence of random variables. 
This turns out to be a relevant modification which may alter the 
basic logarithmic time-dependence of the mean number of records, and it
can induce correlations among the record times, as detected in deviations
of the index of dispersion (\ref{rho}) from unity. 
It is worth noting that evidence for such correlations
can also be found in recent applications of record dynamics in simulations
of complex systems \cite{Sibani03,Anderson04}. An analytic understanding
of the origin of correlations in the models presented here is clearly
an important goal for the near future.

\ack I am grateful to Kavita Jain for her 
contributions in the
early stages of this project, and to Sid Redner for useful correspondence
and discussions. This work was supported by DFG within SFB-TR 12
\textit{Symmetries and universality in mesoscopic systems}.



\end{document}